\documentclass{aastex701}
\usepackage[dvipsnames]{xcolor}

\begin{document}

\title{Study of Chromospheric Global-Scale Flows using Local Correlation Tracking Across the past 10 Solar Cycles}

\author{Qin Li}
\affiliation{Institute for Space Weather Sciences, New Jersey Institute of Technology,
University Heights, Newark, NJ 07102-1982, USA}
\email{ql47@njit.edu}

\author{Yan Xu}
\affiliation{Institute for Space Weather Sciences, New Jersey Institute of Technology,
University Heights, Newark, NJ 07102-1982, USA}
\email{yan.xu@njit.edu}

\author{Keiji Hayashi}
\affiliation{Institute for Space Weather Sciences, New Jersey Institute of Technology,
University Heights, Newark, NJ 07102-1982, USA}
\email{keiji.hayashi@njit.edu}

\author{Chunlei Liang}
\affiliation{Department of Mechanical and Aerospace Engineering,
Clarkson University, Potsdam, NY 13699, USA}
\email{cliang@clarkson.edu}

\author{Haimin Wang}
\affiliation{Institute for Space Weather Sciences, New Jersey Institute of Technology,
University Heights, Newark, NJ 07102-1982, USA}
\email{haimin.wang@njit.edu}

\begin{abstract}

Global-scale solar surface flows, including differential rotation and meridional circulation, provide key observational constraints on the solar dynamo. We investigate these flows using local correlation tracking (LCT) of chromospheric intensity structures in digitized \ion{Ca}{2}~K observations from the Kodaikanal Solar Observatory covering Solar Cycles (SC) 14 to 23 (1907 -- 2007). The most valuable data coverage are from cycles 14 to 19 for two reasons:  (1) it provided record for early cycles that are not commonly available;  (2) it had much better duty cycle than the data of cycles 20-23, therefore the measurements had much less errors.  On the other hand, the information loss in cycles 20-23 are compensated by other modern space observations. The recovered differential rotation shows relatively small cycle-to-cycle variations and remains close to the Howard reference profile. Recurring zonal-flow residual bands at low and middle latitudes are broadly associated with the activity belts during the well-sampled SC~15--19. The meridional flow is predominantly poleward in the earlier record and exhibits larger relative cycle-to-cycle variations and uncertainties. Incomplete coverage of SC~14 and sparse sampling during SC~21--23 limit detailed comparisons to the better-covered SC~15--20. Compared with its immediate successor SC~19, the strongest cycle in the record, SC~18 exhibits a stronger and continuous northern poleward meridional branch, peaking two to three years before its sunspot maximum. SC~19 shows a minor enhancement that peaks earlier prior to its own maximum. An independent comparison using PSPT observations yields a full-period correlation of $r=0.91$ between the LCT and SDO/HMI meridional-flow profiles. These results extend the observational context for solar-cycle-related flows over a century and motivate further refinement through data-driven reconstruction.

\end{abstract}


\section{Introduction} \label{sec:intro}

Differential rotation and meridional flow are important components of the solar dynamo, the mechanism responsible for generating and maintaining the Sun's magnetic field \citep{2005LRSP....2....2C}. In the $\Omega$ effect, differential rotation stretches and twists the Sun's poloidal magnetic field into a toroidal field, amplifying it within the solar convection zone. The meridional flow, in turn, transports magnetic flux from sunspot regions at low latitudes to high latitudes, and eventually to the polar area. During each cycle, the accumulated opposite-polarity flux reverses the polar field near solar maximum, which completes the conversion of the toroidal field into the poloidal field of the following cycle \citep{2005LRSP....2....5S}. The speed of the meridional flow sets the timescale of this surface transport, so slower flows lengthen the ongoing cycle and delay the build-up of the polar field that seeds the following cycle, whereas faster flows shorten it \citep{2014JGRA..119.3316H,2014ApJ...792..142U}. Differential rotation and meridional circulation can jointly modulate solar-cycle amplitude and period. Weaker rotational shear reduces the efficiency of toroidal-field generation, while variations in meridional circulation alter magnetic-flux transport and cycle timing. Their effect on cycle amplitude depends on the balance between advection, diffusion, and magnetic-field amplification \citep{2020LRSP...17....4C, 2008ApJ...673..544Y}.

Differential rotation and meridional flow are measured with three main approaches: direct Doppler measurements, helioseismology, and feature tracking. Doppler measurements determine line-of-sight plasma velocities from spectral-line shifts. Measurements of the weak meridional flow require careful separation from the rotational signal and corrections for observer motion, convective blueshift, and other systematic effects \citep{2010ApJ...725..658U}. Local helioseismic techniques, such as ring-diagram analysis and time--distance helioseismology, infer subsurface flows from changes in oscillation power spectra and acoustic travel times \citep{1993Natur.362..430D,1999ApJ...512..458B}. Feature tracking estimates horizontal motions from the displacements of intensity or magnetic patterns between successive images; magnetic-feature tracking has been widely used to measure large-scale differential rotation and meridional flow \citep{2011ApJ...729...80H}. In addition, other methods are used for studying specific types of flows. For instance, the minimum energy fit method \citep{2004ApJ...612.1181L} has been applied to investigate flows within active regions (ARs), capturing small-scale, localized dynamics. Similarly, machine learning approaches, such as supervised learning, have been developed for detailed studies of flows at the granulation scale \citep{2017A&A...604A..11A}. These methods, while focused on smaller spatial or temporal scales, contribute valuable insights into the interplay between local and global-scale flows.

A key distinction among these methods lies in their sensitivity to different components of the velocity field. Doppler measurements directly measure the line-of-sight component, while feature-tracking techniques predominantly infer motions in the image plane. Helioseismology infers subsurface flows from their effects on solar oscillations. Certain methods, such as the minimum energy fit, use magnetic-field evolution and the induction equation to estimate the photospheric velocity vector under additional constraints \citep{2004ApJ...612.1181L}. Feature tracking itself encompasses a wide range of approaches and is not confined to a single layer of the solar atmosphere: it can be applied to intensitygrams and magnetograms for photospheric flows and to coronal bright points for coronal rotation \citep{2010A&A...520A..29W}. Related flux-modulation methods also use radio features observed at 17~GHz to estimate rotation rates \citep{2009MNRAS.400L..34C}.

Beyond their role as observational diagnostics, measured surface-flow profiles provide constraints on prescribed flows in solar dynamo modeling. In conventional kinematic flux-transport dynamo models, differential rotation shears the poloidal field into a toroidal component, while the assumed meridional circulation transports magnetic flux and influences the period, amplitude, and equatorward migration of the modeled activity belts \citep{2007A&A...474..239J,2008ApJ...673..544Y}. Surface flux-transport models also use observed flow profiles to simulate polar-field evolution \citep{2014ApJ...780....5U}. In dynamic models, magnetic feedback on differential rotation and meridional circulation can produce zonal-flow perturbations associated with torsional oscillations \citep{2006ApJ...647..662R}.

The study of global-scale solar flows has advanced significantly in recent decades through observations from space-based instruments (e.g., SOHO/MDI and SDO/HMI) and ground-based facilities (e.g., GONG and SOLIS). These observations have established the basic structure of differential rotation and its cycle variation: the surface rotation profile is well documented \citep{2000SoPh..191...47B}, and torsional oscillations---bands of faster and slower rotation that migrate with the solar cycle---have been detected in Doppler and magnetic-feature-tracking measurements \citep{1980ApJ...239L..33H,2011ApJ...729...80H} and traced into the solar convection zone through helioseismology \citep{2000ApJ...533L.163H,2002Sci...296..101V}. Surface poleward meridional flows with typical speeds of 10--20~m~s$^{-1}$ are likewise well documented \citep{2010Sci...327.1350H}. Helioseismic studies provide evidence for equatorward return flows at depth, although their depth and cell structure remain uncertain \citep{2013ApJ...774L..29Z,2020Sci...368.1469G}. However, detailed and continuous measurements are concentrated in the past few decades. Systematic reconstructions of torsional oscillations and meridional circulation across multiple earlier cycles remain limited, so the multi-cycle stability of these flows is poorly constrained before the space era.

The century-long record of solar observations, particularly from the Kodaikanal Solar Observatory (KoSO) of the Indian Institute of Astrophysics, offers an opportunity to address this gap. In this study, we analyze 100 years of \ion{Ca}{2} K images from KoSO to derive differential-rotation, torsional-oscillation, and meridional-flow profiles spanning ten solar cycles (SC~14--23). These reconstructed profiles can provide long-baseline observational constraints on the surface components of the prescribed flows used in flux-transport dynamo models, although their extension into the solar interior requires additional assumptions. Section~\ref{sec:method} details the data processing and methodology employed in this work. The results are presented and analyzed in Section~\ref{sec:results}, followed by a discussion of their implications and conclusions in Section~\ref{sec:summary}.

\section{Data processing and methodology} \label{sec:method}

\subsection{Data and Observations}

\begin{figure*}[!t]
\centering\includegraphics[width=6.8in]{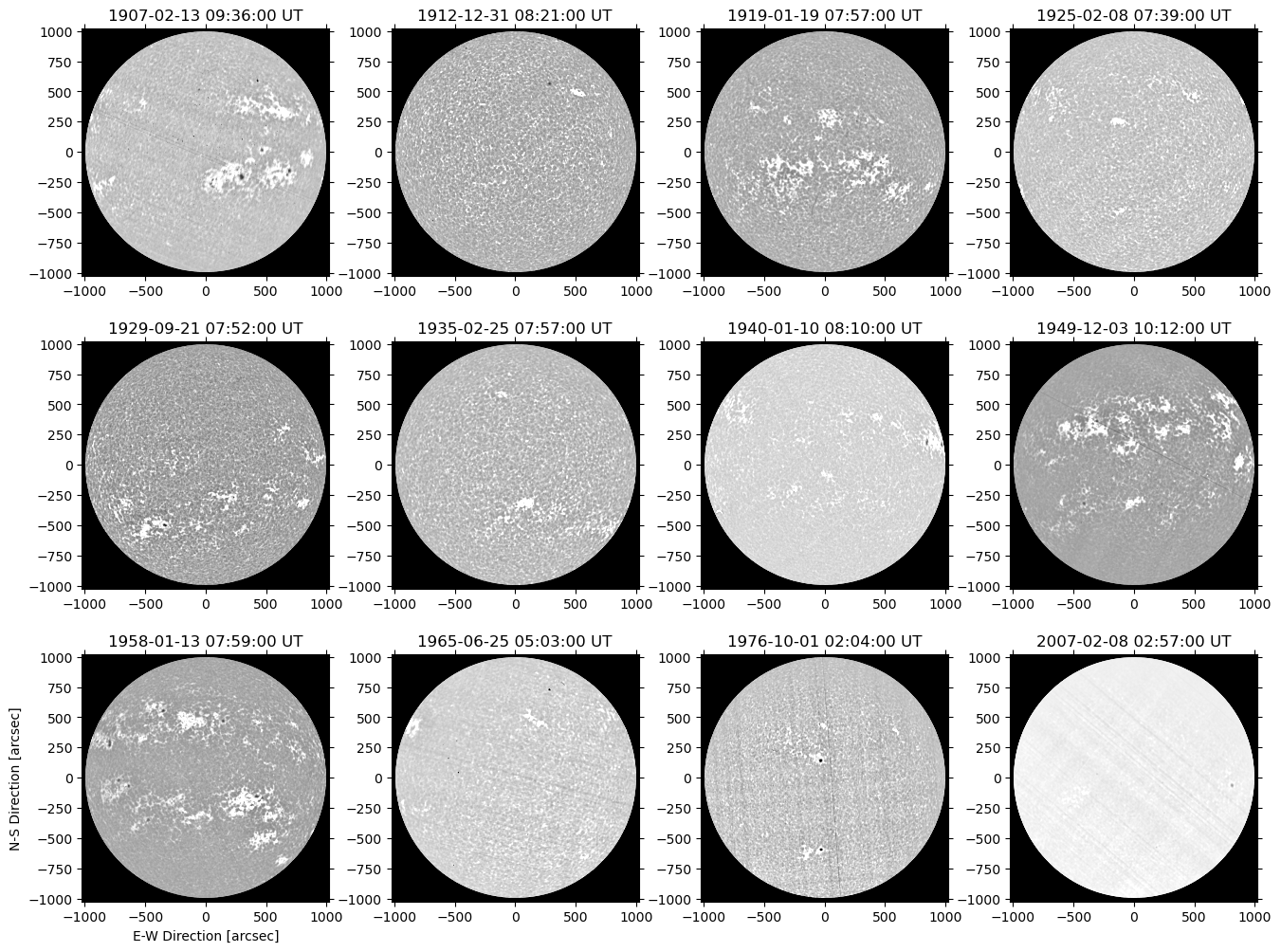}
\caption{Example Ca II K images showcasing chromospheric structures from 1907 to 2007, adaptively scaled for enhanced visibility of features.}
\label{fig:CaK_image}
\end{figure*}

The primary dataset used in this study consists of digitized \ion{Ca}{2} K full-disk images from KoSO. The dataset spans approximately 100 years, from 1907 to 2007, and provides a century-long record of solar chromospheric activity across ten solar cycles (SC~14--23). The \ion{Ca}{2} K spectral line ($\lambda = 393.4$~nm) is a resonance line of singly ionized calcium whose core samples chromospheric layers, while its wings sample photospheric layers. Its intensity responds to magnetic activity, which makes the Ca~II~K line a long-established proxy for solar magnetic activity \citep{2005MmSAI..76.1018O,2014SoPh..289..137P}. Full-disk spectroheliograms reveal chromospheric plages and network structures associated with concentrations of photospheric magnetic flux, although their visibility and height sensitivity depend on the instrumental passband. A few examples of \ion{Ca}{2} K images from this archive are shown in Figure~\ref{fig:CaK_image}.

The KoSO spectroheliograph acquired \ion{Ca}{2} K images on a regular basis throughout the study period, and the resulting archive is one of the longest full-disk Ca~II~K records available \citep{2014SoPh..289..137P}. Its long duration and use of a single instrument make it valuable for century-scale flow measurements, although changes in image quality and calibration must still be considered. The archive has already supported studies of long-term chromospheric feature variation \citep{2017SoPh..292...85P} and a century-long measurement of chromospheric differential rotation \citep{2024ApJ...961...40M}.

The historical \ion{Ca}{2} K images, originally captured on photographic plates, have been digitized with a 4k$\times$4k CCD-based digitizer; on the reduced grid used here, the effective sampling is approximately 4\arcsec\ per pixel. The digitization and subsequent calibration procedures include flat-field correction, density-to-intensity conversion, disk centering, and image rotation to place solar north upward. Detailed descriptions of the digitizer and the Ca~II~K calibration procedures can be found in \citet{2013A&A...550A..19R} and \citet{2014SoPh..289..137P}, respectively. Sunspot areas and positions for 1921--2011 were obtained from the digitized and calibrated KoSO white-light images \citep{2017A&A...601A.106M}.

\begin{figure*}[!t]
\centering
\includegraphics[width=0.9\textwidth]{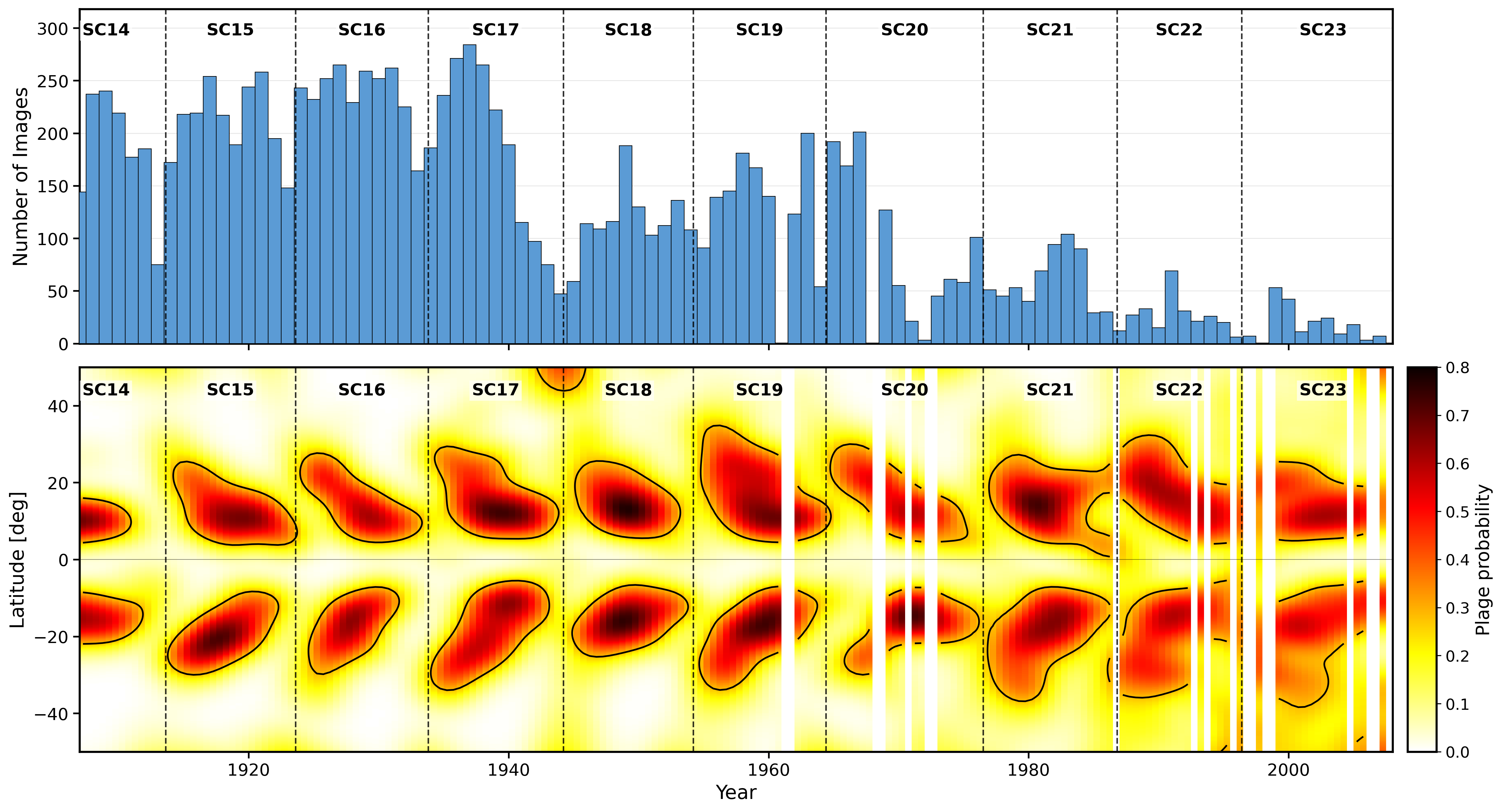}
\caption{Top: distribution of Kodaikanal \ion{Ca}{2} K observations used in this study across Solar Cycles~14--23 (1907--2007), binned per calendar year. Bottom: Ca~II~K plage-probability butterfly diagram constructed from the same archive; the black contours enclose the main plage envelopes. The dominant plages reside within $\sim$40$^{\circ}$ latitude and follow the equatorward migration of activity belts; the white vertical stripes are data gaps.}
\label{fig:sc_histogram}
\end{figure*}

Figure~\ref{fig:sc_histogram} (top) illustrates the data availability of the KoSO \ion{Ca}{2} K sample used in this study: 12,269 daily observations spanning 1907--2007, with counts that vary by more than an order of magnitude from cycle to cycle, from 2401 observations in SC~16 to 195 in SC~23. No usable observations are present in this sample for 1961, 1968, or 1998, and several years after 1970 contain fewer than 20 usable images. The number of available observations affects the quality of the reconstructed flow maps and profiles; in particular, the sparse coverage in SC~22--23 is consistent with the larger uncertainties in the corresponding flow measurements. Figure~\ref{fig:sc_histogram} (bottom) shows the Ca~II~K plage butterfly diagram from the same sample. Plage detection uses an initial smoothing step followed by a threshold of $1.003\times$ the latitude-dependent quiet-Sun level. The resulting map is then smoothed with a gap-preserving Gaussian filter ($\sigma=14$ latitude bins $\approx6^{\circ}$, $\sigma=3$ time bins $\approx1.5$~yr). The black contours mark the 0.28 level of the smoothed plage-occurrence map and are also overlaid on the flow butterfly maps in Section~\ref{sec:time_latitude}. The main plage bands lie within approximately $\pm40^{\circ}$ latitude and provide prominent features for the flow measurements. Their equatorward migration traces the activity bands whose associated flows we analyze below.

\subsection{Methodology}

Local Correlation Tracking (LCT) is employed in this study to derive global-scale flow maps, including differential rotation and meridional flows, from the digitized KoSO \ion{Ca}{2} K images spanning 1907--2007. LCT estimates displacement vectors between local regions in successive images from the peak of their cross-correlation \citep{1988ApJ...333..427N}. We use the implementation of \citet{2011A&A...529A.153V}. Because LCT tracks the apparent motion of intensity structures, its inferred velocities may differ from the underlying plasma velocities. Tests with simulated granulation demonstrate an amplitude underestimate that depends on the tracking parameters \citep{2013A&A...555A.136V}. We therefore apply an empirical amplitude calibration, as described in Section~\ref{sec:calibration}.

\begin{figure*}[!t]
\centering
\includegraphics[width=1\textwidth]{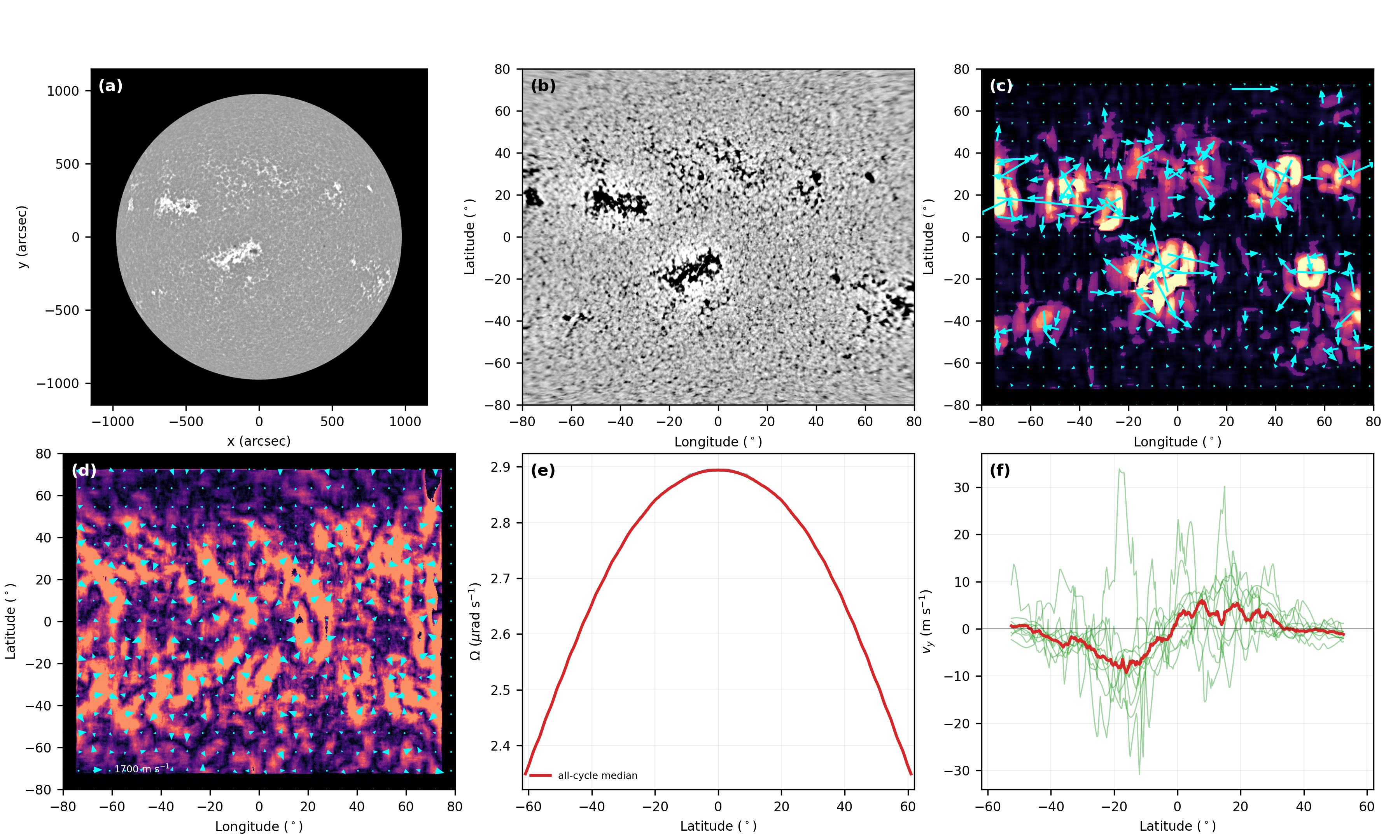}
\caption{Demonstration of the pipeline from a KoSO \ion{Ca}{2} K full-disk observation (1927 January 1) to the final flow products: (a) raw full-disk spectroheliogram, (b) Carrington heliographic map of the processed observation, (c) LCT displacement/velocity field from a 1-day image pair, (d) cycle-mean LCT velocity field, (e) all-cycle median sidereal differential rotation $\Omega(\lambda)$, and (f)cyclic meridional flow profiles $v_y(\lambda)$ (green curves) with the all-cycle median (red curve).}
\label{fig:sc_workflow}
\end{figure*}

For the century-long \ion{Ca}{2} K time series, we use a $40\times40$-pixel cross-correlation subfield and a Gaussian apodization window with a nominal full width at half maximum (FWHM) of 120~Mm at the equator. This broad window combines information from multiple chromospheric structures to emphasize large-scale motions and reduce sensitivity to small-scale intensity fluctuations. Differential-rotation compensation reduces the displacement between image pairs before the LCT calculation and facilitates detection of the residual motion within the search window.

The processing workflow comprises five steps: pre-processing, heliographic mapping, flow compensation, the LCT procedure, and velocity conversion. In pre-processing, each digitized \ion{Ca}{2} K image is first resampled to $512\times512$ pixels for computational efficiency. Zero-valued pixels inside the disk are replaced by the regional mean, and an iterated Gaussian high-pass filter (30 passes at $\sigma=5$ pixels) suppresses large-scale intensity gradients. The residual image is then scaled to a common dynamic range through a disk-masked, $\sigma$-clipped normalization. These steps reduce background variations and standardize the intensity range across plates of varying photographic quality. Each processed image is projected onto a Lambert equal-area heliographic grid with the observation date and solar ephemeris, including the $P$ angle and $B_0$. After edge removal and resampling, the final maps contain $429\times607$ pixels and span approximately $\pm61^{\circ}$ in latitude. 

Prior to the LCT calculation, the second image of each pair is shifted in longitude at each latitude by an integer number of pixels based on a prescribed differential-rotation profile \citep{1990SoPh..130..295H}. The adopted sidereal rotation rate ranges from 14.33$^{\circ}$~day$^{-1}$ at the equator to 10.38$^{\circ}$~day$^{-1}$ at the poles. This compensation reduces the residual displacement between the two images. The LCT calculation is performed in the Fourier domain: Gaussian-apodized $40\times40$-pixel subfields are cross-correlated, and the integer correlation peak is located within a circular search radius of 10 pixels. A five-point parabolic fit refines the peak position to sub-pixel accuracy in both coordinates, and a second correlation around the refined displacement provides a consistency check to reject spurious detections. Image pairs with time separations outside 0.5--2.5 days are discarded. The measured displacements are converted to velocities using the pair time separation and the spatial scale of the projected grid.

The per-pair products are reduced to one longitude-averaged profile per observation date. For the zonal component, the pipeline applies a resistant mean over longitude and rejects velocities that deviate from the median by more than $3\sigma$, where the deviation scale is estimated from the median absolute deviation. For the meridional component, the displacement cube is averaged over longitude after removal of the 150 outermost columns on each side of the map. The pair time separation is included in the conversion from displacement to velocity. From these daily velocity profiles, we construct cycle-mean profiles and time--latitude diagrams. Cycle averages use a Tukey biweight estimator \citep{1990AJ....100...32B} to reduce the influence of outlying daily profiles.


\section{Results} \label{sec:results}

\subsection{Amplitude calibration against SDO/HMI} \label{sec:calibration}

\begin{figure*}[!t]
\centering\includegraphics[width=6.8in]{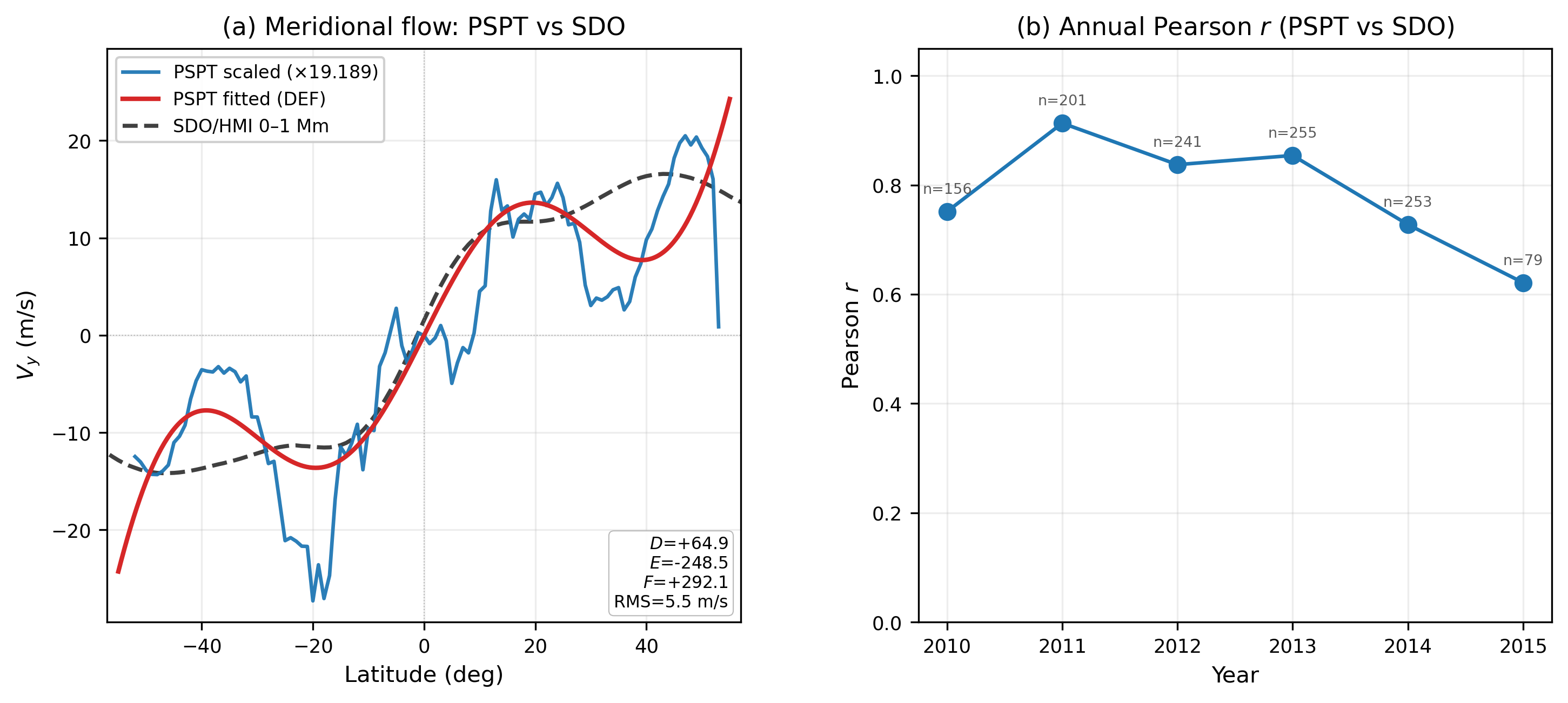}
\caption{Comparison of meridional-flow profiles from PSPT \ion{Ca}{2}~K LCT and SDO/HMI time-distance helioseismology over the 2010--2015 overlap period (1185 matched days). \emph{Left:} the blue solid curve shows the PSPT LCT profile after application of the fitted amplitude scale factor ($s=19.189$), the red solid curve shows its three-term antisymmetric fit, and the black dashed curve shows the HMI profile at a depth of $0$--$1$~Mm. Positive velocities indicate northward motion. \emph{Right:} annual Pearson correlation coefficients between the PSPT and HMI profiles constructed from matched observing days; the number of matched days is annotated for each year.}
\label{fig:pspt_sdo}
\end{figure*}

Because LCT-derived velocities can underestimate the underlying plasma velocities, we first establish an empirical amplitude scale using modern observations. We measured meridional flows with the same LCT pipeline in \ion{Ca}{2}~K full-disk observations from the Precision Solar Photometric Telescope (PSPT), whose overlap with SDO/HMI covers 2010 May 1 to 2015 June 8. Figure~\ref{fig:pspt_sdo} compares the PSPT profile with the HMI time--distance profile at a depth of $0$--$1$~Mm over 1185 matched days. After application of a single amplitude scale factor of $s=19.2$, the full-period profiles have a Pearson correlation of $r=0.91$ and a reported RMS difference of $5.5~\mathrm{m\,s^{-1}}$. Both profiles recover the broad pattern of poleward flow in the two hemispheres, and their principal zero crossings occur close to the equator. The PSPT profile reaches approximately $10$--$15~\mathrm{m\,s^{-1}}$ at northern latitudes of $15^{\circ}$--$25^{\circ}$, comparable to HMI. However, it exhibits substantially stronger latitude-dependent fluctuations, particularly near $20^{\circ}$ south and $30^{\circ}$--$40^{\circ}$ north. The agreement therefore concerns the large-scale latitude dependence rather than all local extrema or amplitudes.

\begin{figure*}[!t]
\centering\includegraphics[width=6.8in]{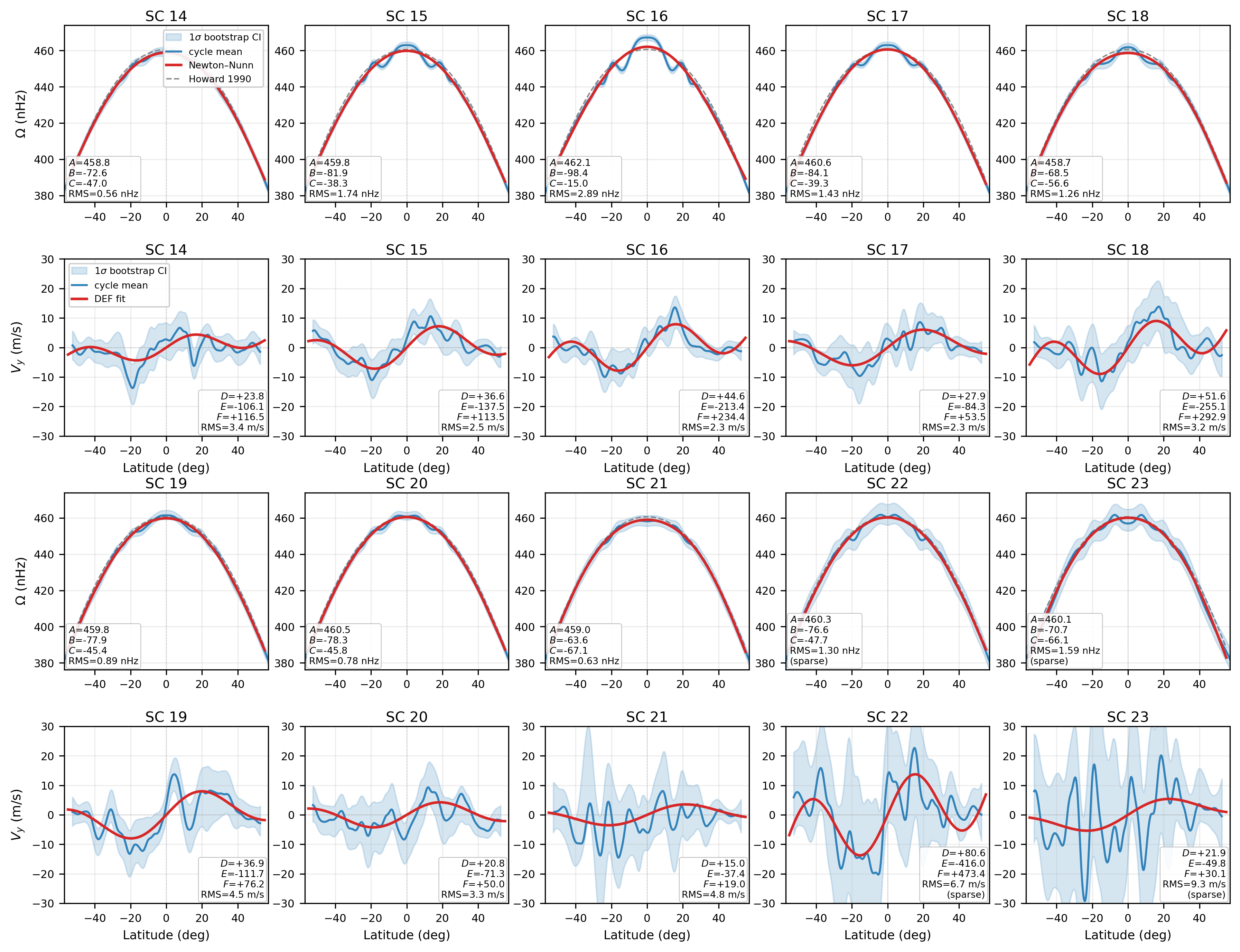}
\caption{Cycle-mean surface differential rotation (upper panel of each pair) and meridional flow (lower panel of each pair) for Solar Cycles~14--23, derived from KoSO local correlation tracking measurements. The blue curve is the cycle mean and the shaded band its $1\sigma$ bootstrap confidence interval; the solid red curves show the three-term fits, and the dashed black curve the \citet{1990SoPh..130..295H} reference profile adopted for rotation compensation. Positive $V_y$ is northward.}
\label{fig:cycle_mean_DR_meridional}
\end{figure*}

Some of these fluctuations may reflect converging flows toward active regions superimposed on the background poleward circulation. Such inflows enhance the poleward velocity on the equatorward side of an activity belt and reduce it on the poleward side, which can produce local maxima and minima in the meridional-flow profile \citep{2023ApJ...950...63M}. This interpretation is particularly relevant to \ion{Ca}{2}~K feature tracking, for which bright plages and network structures provide prominent tracers. The inferred profile may consequently give greater weight to strong magnetic concentrations such as active regions than a spatially averaged plasma-flow measurement. At higher latitudes, the lower abundance of plages may reduce tracking sensitivity and increase uncertainty, although network structures remain available as tracers. This sampling effect could contribute to local velocity reductions and discrepancies with HMI, but its systematic nature may enable future bias correction through machine learning.

The annual Pearson correlations range from approximately $0.73$ to $0.91$ during 2010--2014 and decrease to $0.62$ in 2015. The latter year includes only 79 matched days through June 8, so its lower correlation may partly reflect the shorter sampling interval. These results support persistent agreement in the meridional profiles across years. Please note adopt the scale factor for the flow profiles throughout this work, with the assumption that the empirical calibration remains applicable to the historical observations despite differences in image quality and instruments. 

\subsection{Cycle-mean rotation and meridional flow} \label{sec:cycle_mean}

Figure~\ref{fig:cycle_mean_DR_meridional} presents the cycle-averaged surface differential rotation (upper panel of each pair) and meridional flow (lower panel of each pair) for Solar Cycles~14--23, obtained from KoSO local correlation tracking measurements spanning 1907--2007. Overall, the rotation profiles are symmetrized about the equator, while the meridional profiles are antisymmetrized. The meridional profiles are fitted with the antisymmetric form adopted by \citet{2014ApJ...792..142U},
\begin{equation}
v_\theta(\lambda)
=
\left[D\sin\lambda+E\sin^3\lambda+F\sin^5\lambda\right]\cos\lambda,
\label{eq:meridional_fit}
\end{equation}
where $\lambda$ is latitude and positive velocity denotes northward motion. The rotation frequency is fitted with the standard latitude-dependent form
\begin{equation}
\nu(\lambda)\equiv\frac{\Omega(\lambda)}{2\pi}
=
A+B\sin^2\lambda+C\sin^4\lambda,
\end{equation}
where $A$, $B$, and $C$ are expressed in nHz.  Before fitting, the cycle-mean meridional profiles are Gaussian-smoothed along latitude ($\sigma=5$ latitude bins $\approx1.4^{\circ}$). All fits are restricted to $|\lambda|\leq55^\circ$.

The recovered differential rotation profiles show relatively small cycle-to-cycle differences. Restricting the comparison to the six well-sampled cycles SC~15--SC~20 (Figure~\ref{fig:sc_histogram}, top), the fitted equatorial sidereal rotation frequency, $A$, ranges from $458.68$ to $462.08~\mathrm{nHz}$, while $B$ and $C$ range from $-98.42$ to $-68.55~\mathrm{nHz}$ and from $-56.58$ to $-14.95~\mathrm{nHz}$, respectively. The residual RMS values are $0.8$--$2.9~\mathrm{nHz}$ across the fitting interval. Because the Howard-form rotation profile of \citealp{1990SoPh..130..295H} is used for rotation compensation and subsequent restoration, close agreement with this reference alone does not establish the accuracy of the independently recovered residual motion; the fitted high-order terms depart from it by a few $\mathrm{nHz}$ at the largest latitudes sampled. We therefore focus on departures from the mean rotation profile in Section~\ref{sec:time_latitude}.

The meridional profiles exhibit pronounced cycle-to-cycle differences in amplitude and shape. In the displayed fits, the dominant low-latitude branches are poleward and generally reach their local maxima near $16^\circ$--$20^\circ$. Restricting attention to the six well-sampled cycles SC~15--SC~20, these peak speeds span $4.2$ to $9.0~\mathrm{m\,s^{-1}}$, with SC~18 the strongest and SC~20 the weakest. These comparisons refer specifically to the fits displayed in the figure. Antisymmetry follows from the adopted fit form and does not mean hemispheric symmetry in the original measurements. The residual scatter also varies among cycles, reaching $4.5~\mathrm{m\,s^{-1}}$ in SC~19 within the well-sampled set.

We exclude SC~14, SC~21, SC~22, and SC~23 from the quantitative comparisons in this section. SC~14 is truncated at both ends of the KoSO record, so its fit describes a partial, mostly rising-phase segment rather than a cycle mean; SC~21, SC~22, and SC~23 are limited instead by low image counts (639, 260, and 195 usable frames, respectively, against 1028--2401 for SC~15--SC~20). The apparent extremes of the full ten-cycle set are drawn precisely from these four cycles, which is why we do not treat them as measurements of the circulation. Figure~\ref{fig:sc_histogram} (top) shows the coverage underlying both effects. Truncation removes phases of the cycle outright, whereas sparse sampling retains the full phase range but weights it unevenly and inflates the formal uncertainty; both weaken any statement about where the meridional flow is strongest or weakest. A further caveat applies to the fits across all cycles: the DEF basis is not orthogonal over the restricted latitude interval, so $D$, $E$, and $F$ are strongly coupled and $E\sin^3\lambda$ and $F\sin^5\lambda$ nearly cancel over the range where data exist. Individual coefficients therefore cannot be read as independent changes in the circulation, and comparisons of the complete fitted profiles provide the most direct measures of cycle-to-cycle differences.

For comparison, \citet{2014ApJ...792..142U} adopted an average north--south antisymmetric meridional-flow profile of the form in Eq.~(\ref{eq:meridional_fit}), with coefficients $D=24$, $E=16$, and $F=-37~\mathrm{m\,s^{-1}}$. For the well-sampled cycles SC~15--SC~20, our tabulated coefficients have comparable leading amplitudes but high-order terms that are both larger in magnitude and of opposite sign in $E$: $D$ ranges from $20.77$ to $51.59~\mathrm{m\,s^{-1}}$, $E$ from $-255.06$ to $-71.29~\mathrm{m\,s^{-1}}$, and $F$ from $50.03$ to $292.88~\mathrm{m\,s^{-1}}$. $E$ is negative in every one of these cycles, whereas the reference has $E>0$; $F$ is positive in all six as well, opposite in sign to the reference. Our solutions differ from the smooth reference because the high-order terms are large and strongly coupled, so their individual values cannot be interpreted directly as independent changes in the circulation. Comparisons of the complete fitted profiles and peak speeds provide more direct measures of cycle-to-cycle differences.

\begin{table*}[t]
\centering
\caption{Cycle-averaged differential-rotation and flow parameters derived from LCT measurements for Solar Cycles~14--23.
The differential-rotation coefficients $(A,B,C)$ are from Newton--Nunn fits of the symmetrized cycle-mean profiles,
$\nu(\lambda)=A+B\sin^2\lambda+C\sin^4\lambda$ \citep{1951MNRAS..111..413N}, while meridional flow profiles are fitted using Eq.~(\ref{eq:meridional_fit}),
$v_\theta(\lambda) = \left[D\sin\lambda + E\sin^3\lambda + F\sin^5\lambda\right]\cos\lambda$ \citep{2014ApJ...792..142U},
where $\lambda$ is heliographic latitude.
}
\label{tab:cycle_flow_params}
\begin{tabular}{c ccc ccc}
\hline
\hline
& \multicolumn{3}{c}{\textbf{Differential Rotation}} &
\multicolumn{3}{c}{\textbf{Meridional Flow}} \\
\cline{2-4} \cline{5-7}
\textbf{Solar Cycle} &
\boldmath$A$ \textbf{(nHz)} &
\boldmath$B$ \textbf{(nHz)} &
\boldmath$C$ \textbf{(nHz)} &
\boldmath$D$ \textbf{(m s$^{-1}$)} &
\boldmath$E$ \textbf{(m s$^{-1}$)} &
\boldmath$F$ \textbf{(m s$^{-1}$)} \\
\hline
SC 14 & 459.29 & $-73.57$ & $-46.37$ & 23.83 & $-106.07$ & 116.53 \\
SC 15 & 460.42 & $-81.71$ & $-39.85$ & 36.58 & $-137.46$ & 113.46 \\
SC 16 & 462.25 & $-94.23$ & $-22.22$ & 44.61 & $-213.41$ & 234.44 \\
SC 17 & 461.31 & $-82.55$ & $-42.71$ & 27.91 & $-84.30$ & 53.52 \\
SC 18 & 459.22 & $-66.10$ & $-62.96$ & 51.59 & $-255.06$ & 292.88 \\
SC 19 & 460.12 & $-67.54$ & $-62.34$ & 36.85 & $-111.72$ & 76.19 \\
SC 20 & 460.80 & $-74.54$ & $-51.47$ & 20.77 & $-71.29$ & 50.03 \\
SC 21 & 459.75 & $-61.55$ & $-72.33$ & 15.04 & $-37.43$ & 19.01 \\
SC 22 & 461.73 & $-68.19$ & $-64.80$ & 80.63 & $-415.96$ & 473.35 \\
SC 23 & 460.67 & $-64.49$ & $-77.41$ & 21.87 & $-49.78$ & 30.08 \\
\hline
\hline
\end{tabular}
\end{table*}

\subsection{Time--latitude evolution} \label{sec:time_latitude}

\begin{figure*}[!t]
\centering\includegraphics[width=1.0\textwidth]{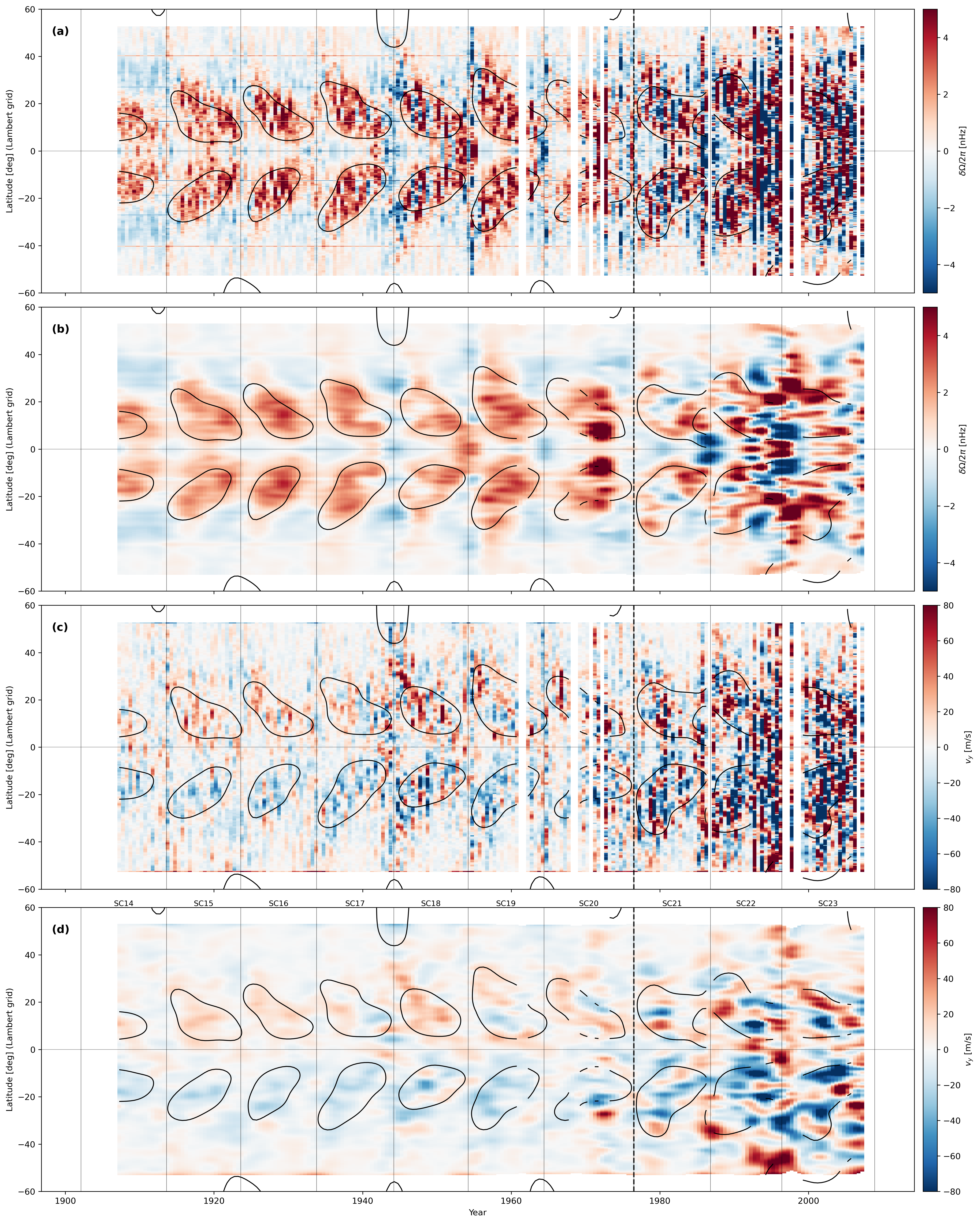}
\caption{Time--latitude (butterfly) maps of the torsional oscillation and meridional flow, overlaid with the Ca~\textsc{ii}~K plage envelope (black contours). \emph{Panels (a)} and \emph{(c)}: zonal-flow residual $\delta\Omega/2\pi$ (color scale $\pm5$~nHz) and meridional flow $v_y$ (color scale $\pm80$~\mbox{m\,s$^{-1}$}) from the binned maps. \emph{Panels (b)} and \emph{(d)}: the same with smoothing. The vertical dashed line at 1976 marks the start of cycle~21, and white regions indicate missing data.}
\label{fig:torsional_meridional}
\end{figure*}

Figure~\ref{fig:torsional_meridional} compares the time--latitude maps of the zonal-flow residuals and meridional velocities, with the \ion{Ca}{2}~K plage envelope overlaid. Panels~(a) and (c) contain the unsmoothed maps of the zonal residual and meridional velocity, respectively, constructed from 0.5-year quality-control-weighted biweight bins. Panels~(b) and (d) apply a two-dimensional Gaussian filter with $\sigma=2$ latitude rows and $\sigma=2$ time bins ($1$~yr). The unsmoothed maps contain substantial small-scale scatter and vertical striping, especially in the later decades. These features are consistent with residual measurement errors and uneven temporal sampling. The vertical dashed line at 1976.5 marks the approximate beginning of SC~21, beyond which image counts are low. The large amplitudes during this sparsely sampled period should therefore be interpreted with caution. Smoothing makes the broad structures easier to identify, although the temporal filter has an FWHM of approximately $2.35$~yr and can broaden or merge neighboring features.

The most evident recurring structure in the zonal maps is a sequence of bands at low and middle latitudes that broadly follows the equatorward migration of the plage belts. These bands are particularly apparent during SC~15--19, when positive residuals overlap with the plage envelope and negative residuals occur preferentially on its poleward side. The recurring association between these bands and the activity belts suggests a cycle-related component in the zonal signal.

More generally, positive meridional velocities in the northern activity belt and negative velocities in the southern activity belt indicate predominantly poleward motion. This pattern broadly follows the latitude range occupied by the plages, which provide prominent intensity structures for feature tracking. Consequently, the apparent association between flow amplitude and plage occurrence may reflect both physical flow variations and changes in tracer sensitivity. Local departures from the broad poleward pattern could include contributions from active-region inflows, but the present maps do not isolate these contributions from the background flows.

The time--latitude maps are qualitatively consistent with helioseismic evidence for solar-cycle variations in large-scale flows. \citet{2025SoPh..300..149K} identified bands of zonal deceleration and meridional convergence associated with magnetic activity, together with fast zonal bands that precede them at low and middle latitudes. Their analysis also identified a possible early flow signature of SC~26. Though Their measurements resolve subsurface depth dependence, whereas our \ion{Ca}{2}~K analysis tracks apparent motions of intensity structures on chromosphere.

We examine SC~18 and SC~19 together to search for possible flow signatures preceding the exceptionally strong SC~19. In the zonal maps, both cycles show low-latitude positive bands broadly associated with the plage envelope, with negative residuals preferentially on its poleward side. An additional positive feature appears near the equator toward the end of SC~18, approaching the transition to SC~19. The meridional flow also shows a clearer contrast between the two cycles, particularly at northern middle latitudes, where the RMS velocity during SC~18 exceeds that of the preceding three cycles, with $3.2~\mathrm{m,s^{-1}}$. The poleward flows appears more pronounced during SC~18, whereas it becomes weaker again during SC~19. The contrast is particularly apparent during the rising phases, with the SC~19 enhancement occurring earlier relative to its maximum, approximately three years before maximum.

These maps extend the investigation of activity-associated surface-flow patterns across SC~14--23 and complement the century-long differential-rotation measurements \citep{2024ApJ...961...40M}. The SC~18--19 comparison highlights potentially informative variations, but does not yet establish a flow precursor to the exceptional strength of SC~19. Quantitative interpretation of the zonal amplitudes remains subject to the calibration limitations, while interpretation of both components requires consideration of temporal coverage and tracer availability.

\begin{figure*}[!t]
\centering\includegraphics[width=1.0\textwidth]{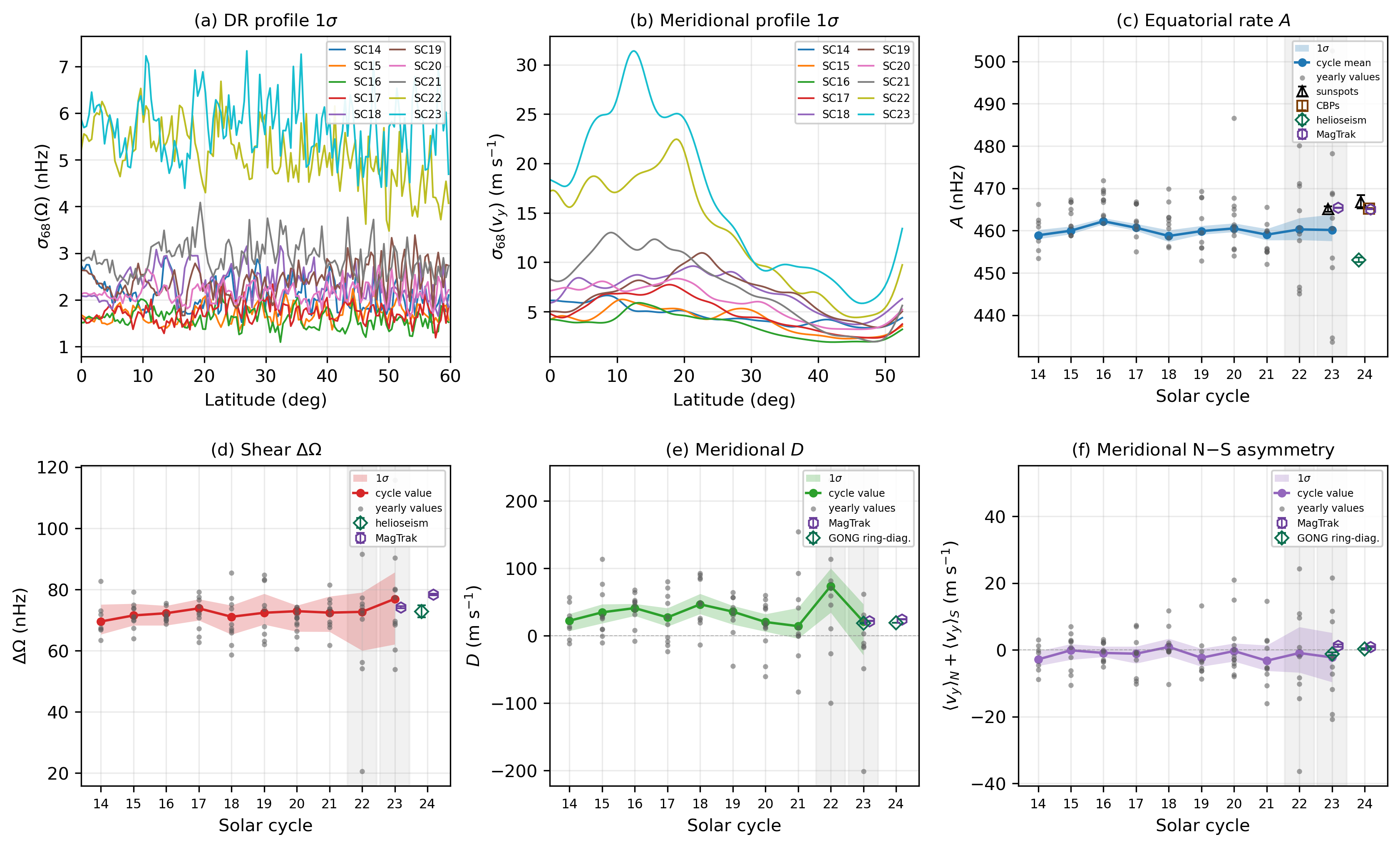}
\caption{Cycle-mean surface-flows and bootstrap uncertainties for Solar Cycles~(SC)~14--23.
(a)~Latitude-dependent $1\sigma$ half-width $\sigma_{68}(\Omega)$ of the cycle-mean differential-rotation profile.
(b)~Same for the meridional profile $\sigma_{68}(v_y)$.
(c)~Equatorial sidereal rate $A=\Omega(0)$.
(d)~Latitudinal shear $\Delta\Omega=\Omega(0)-\Omega(55^{\circ})$.
(e)~Coefficient $D$ of the meridional fit.
(f)~North--South meridional asymmetry.
Colored curves show cycle means, with shaded bands indicating $1\sigma$ bootstrap uncertainties; gray dots show year-level estimates.
Markers for SC~23--24 are independent comparisons for each quantity: sunspot tracking \citep{2017A&A...606A..72P}, coronal bright points \citep{2024ApJ...972...46S} for rotation, HMI global modes \citep{2018SoPh..293...29L}, HMI ring-diagram shear \citep{2026arXiv260819438R}, MagTrak network LCT \citep{2021ApJ...917..100M}, and GONG MRRFP ring diagrams \citep{2018SoPh..293..145K}.
}
\label{fig:uncertainty}
\end{figure*}

\subsection{Uncertainty analysis}
\label{sec:uncertainty}

Figure~\ref{fig:uncertainty} summarizes the uncertainties associated with the cycle-mean profiles. Panels~(a) and~(b) show the latitude dependence of the central $68\%$ bootstrap interval half-widths, defined as $\sigma_{68}=(P_{84}-P_{16})/2$, where $P_{16}$ and $P_{84}$ are the corresponding bootstrap percentiles. This measure approximates a $1\sigma$ uncertainty for a Gaussian distribution. Across all the cycles, $\sigma_{68}(\nu)$ is typically a few~$\mathrm{nHz}$ from the equator to approximately $55^{\circ}$, while $\sigma_{68}(v_y)$ remains of order several~$\mathrm{m\,s^{-1}}$. SC~22 and especially SC~23 show substantially larger uncertainties, with rotation half-widths of approximately $4$--$7~\mathrm{nHz}$ and meridional half-widths exceeding approximately $20$--$30~\mathrm{m\,s^{-1}}$ near mid-latitudes. This increase coincides with reduced observational coverage.

Panel~(c) shows comparatively small cycle-to-cycle differences in the calibrated equatorial sidereal rotation frequency: cycle means cluster near $A\simeq459$--$462~\mathrm{nHz}$, with year-to-year scatter of a few~$\mathrm{nHz}$ across the cycles and larger scatter and bootstrap intervals in SC~22--23. The comparison values shown for magnetic-network tracking, sunspots, and coronal bright points generally greater than our \ion{Ca}{2}~K pattern-tracking estimates \citep{2021ApJ...917..100M,2003SoPh..212...23J,2024ApJ...972...46S}, whereas the selected HMI global-mode comparison lies below them \citep{2018SoPh..293...29L}. These differences provide context for the recovered rotation rates, but should not be interpreted as a universal ordering of tracers. The measurements differ in observing period, tracer properties, spatial sensitivity, and analysis method. 

Panel~(d) shows similarly modest variations in the latitudinal shear, defined as $\Delta\nu=\nu(0)-\nu(55^\circ)$. Values are approximately $70$--$74~\mathrm{nHz}$ across the cycles and rise to approximately $77~\mathrm{nHz}$ in SC~23, where the bootstrap uncertainty is also substantially larger. The apparent increase should therefore be assessed relative to its uncertainty before being interpreted as a physical trend. Comparisons with magnetic-network tracking and HMI ring-diagram measurements \citep{2021ApJ...917..100M,2026arXiv260819438R} are present.

Panels~(e) and~(f) show the meridional-fit coefficient D and meridional N-S asymmetry across the cycles. The leading meridional-fit coefficient $D$ spans approximately $15$--$81~\mathrm{m\,s^{-1}}$ across the cycles shown, with the largest estimate in SC~22 ($D\sim81~\mathrm{m\,s^{-1}}$) accompanied by a wide bootstrap interval. Yearly estimates scatter by tens to $\gtrsim100~\mathrm{m\,s^{-1}}$. Because $D$ is coupled to the higher-order coefficients $E$ and $F$, its variation does not directly measure variation in the peak poleward speed. Overall, the meridional-flow estimates are broadly compatible with the MagTrak and GONG measurements \citep{2021ApJ...917..100M,2018SoPh..293..145K}, although a complete comparisons require complete velocity profiles or coefficients derived using the same fitting function and latitude range.

The north--south asymmetry estimates remain near zero for most cycles, with the largest negative value occurring in SC~19. A negative value of $(v_y){\rm N}+(v_y){\rm S}$ at matched latitudes indicates a stronger southern poleward component when both hemispheric flows are poleward. The significance of the SC~19 departure depends on its bootstrap interval. Modern MagTrak and GONG measurements are included for comparison and broadly agree with our estimates within the uncertainties.

\section{Summary and Discussion} \label{sec:summary}

We applied LCT to the digitized Kodaikanal \ion{Ca}{2}~K full-disk archive to reconstruct chromospheric global-scale horizontal flows from 1907 to 2007, spanning Solar Cycles~14--23. For the first time, this work provides a century-long record of differential rotation, zonal-flow residuals, and meridional flow profiles with the uncertainties.

The cycle-mean differential rotation is stable over the record. The equatorial sidereal rate clusters near $459$--$462~\mathrm{nHz}$, and the equator-to-$55^\circ$ shear is $70$--$74~\mathrm{nHz}$ across the well-sampled cycles. The larger apparent change and substantially broader uncertainty in SC~22--23 coincide with the sharp decline in effective sampling.

The time--latitude maps contain recurring zonal bands at low and middle latitudes that broadly follow the equatorward migration of the plage belts, most clearly during SC~15--19. Their presence in both the unsmoothed and smoothed maps supports a large-scale cycle-related component in the recovered signal. The maps do not, however, establish a common migration rate, a fixed phase relative to the activity belts, or a systematic dependence of band amplitude on cycle strength. The calibrated zonal amplitudes also remain provisional because the PSPT--HMI comparison does not independently constrain the zonal scale.

The meridional flow is predominantly poleward during much of the earlier record, with cycle-dependent changes in profile amplitude and shape. Its cycle means and annual values are more variable than those of the rotation, although the north--south asymmetry remains close to zero for most cycles; for cycles that generate no yearly flow products, only cycle-mean values are available. The meridional-flow results become markedly less certain in SC~22--23, when temporal coverage is sparse, and the cycle-to-cycle comparisons are further compromised in SC~14 by the truncation of that cycle in the KoSO record. We therefore base the meridional amplitude comparisons on SC~15--SC~20, where the fitted low-latitude peak speeds span $4.2$--$9.0~\mathrm{m\,s^{-1}}$ and the spread is comparable to the per-cycle bootstrap uncertainties. A full-period comparison between PSPT LCT and SDO/HMI gives $r=0.91$ after application of a single amplitude scale factor, supporting recovery of the broad latitudinal profile. This comparison does not establish that the scale factor is stable across time or transferable without bias to the historical KoSO plates.

The activity-associated organization of the zonal and meridional maps is qualitatively consistent with modern Doppler and helioseismic measurements, but it does not by itself demonstrate magnetic back-reaction. \ion{Ca}{2}~K plages and enhanced network are prominent magnetic tracers, so changes in tracer distribution and sensitivity can contribute to the apparent flow variations. The reconstructed profiles therefore provide useful long-baseline constraints for flux-transport and dynamical models, provided that model comparisons incorporate the observational coverage, temporal smoothing, and tracer response rather than treating every local feature as a plasma-flow measurement.

Several limitations remain. LCT measures the apparent motion of intensity structures and can underestimate plasma velocities; historical image quality, feature evolution, and calibration artifacts add further uncertainty. The modern amplitude calibration is transferred from PSPT to KoSO without an independent instrument-transfer test, and the temporal smoothing can broaden or merge features on approximately $2.35$-yr scales. Controlled synthetic-shift and degradation experiments, comparisons with overlapping archives, and profile measurements with and without active-region masks will be needed to separate physical flow changes from measurement and tracer effects. Such tests would also permit firmer comparisons with cycle amplitude, hemispheric phase lags, and polar-field reversal timing.

\begin{acknowledgments}
This work uses the digitized \ion{Ca}{2}~K archive of the Kodaikanal Solar Observatory, operated by the Indian Institute of Astrophysics (IIA), Bengaluru, India. We thank the IIA staff for digitizing and preserving the historical observations. 
\end{acknowledgments}

% ...

\bibliography{references}{}

@string{apj     = "The Astrophysical Journal"}

@string{apjl     = "The Astrophysical Journal Letters"}

@string{aap      = "Astronomy $\&$ Astrophysics"}

@string{solphys     = "Solar Physics"}

@string{mnras      = "Monthly Notices of the Royal Astronomical Society"}

@ARTICLE{2021ApJ...917..100M,
       author = {{Mahajan}, Sushant S. and {Hathaway}, David H. and {Mu{\~n}oz-Jaramillo}, Andr{\'e}s and {Martens}, Petrus C.},
        title = "{Improved Measurements of the Sun's Meridional Flow and Torsional Oscillation from Correlation Tracking on MDI and HMI Magnetograms}",
      journal = apj,
         year = 2021,
        month = aug,
       volume = {917},
       number = {2},
          eid = {100},
        pages = {100},
          doi = {10.3847/1538-4357/ac0a80},
archivePrefix = {arXiv},
       eprint = {2107.07731},
 primaryClass = {astro-ph.SR},
       adsurl = {https://ui.adsabs.harvard.edu/abs/2021ApJ...917..100M}
}

@ARTICLE{2000ApJ...533L.163H,
       author = {{Howe}, R. and {Christensen-Dalsgaard}, J. and {Hill}, F. and {Komm}, R.~W. and {Larsen}, R.~M. and {Schou}, J. and {Thompson}, M.~J. and {Toomre}, J.},
        title = "{Deeply Penetrating Banded Zonal Flows in the Solar Convection Zone}",
      journal = {\apjl},
         year = 2000,
        month = apr,
       volume = {533},
       number = {2},
        pages = {L163-L166},
          doi = {10.1086/312623},
archivePrefix = {arXiv},
       eprint = {astro-ph/0003121},
 primaryClass = {astro-ph},
       adsurl = {https://ui.adsabs.harvard.edu/abs/2000ApJ...533L.163H}
}

@ARTICLE{2002Sci...296..101V,
       author = {{Vorontsov}, S.~V. and {Christensen-Dalsgaard}, J. and {Schou}, J. and {Strakhov}, V.~N. and {Thompson}, M.~J.},
        title = "{Helioseismic Measurement of Solar Torsional Oscillations}",
      journal = {Science},
         year = 2002,
        month = apr,
       volume = {296},
       number = {5565},
        pages = {101-103},
          doi = {10.1126/science.1069190},
       adsurl = {https://ui.adsabs.harvard.edu/abs/2002Sci...296..101V}
}

@ARTICLE{2006ApJ...647..662R,
       author = {{Rempel}, Matthias},
        title = "{Flux-Transport Dynamos with Lorentz Force Feedback on Differential Rotation and Meridional Flow: Saturation Mechanism and Torsional Oscillations}",
      journal = {\apj},
         year = 2006,
        month = aug,
       volume = {647},
       number = {1},
        pages = {662-675},
          doi = {10.1086/505170},
archivePrefix = {arXiv},
       eprint = {astro-ph/0604446},
 primaryClass = {astro-ph},
       adsurl = {https://ui.adsabs.harvard.edu/abs/2006ApJ...647..662R}
}

@ARTICLE{2008ApJ...673..544Y,
       author = {{Yeates}, Anthony R. and {Nandy}, Dibyendu and {Mackay}, Duncan H.},
        title = "{Exploring the Physical Basis of Solar Cycle Predictions: Flux Transport Dynamics and Persistence of Memory in Advection- versus Diffusion-dominated Solar Convection Zones}",
      journal = {\apj},
         year = 2008,
        month = jan,
       volume = {673},
       number = {1},
        pages = {544-556},
          doi = {10.1086/524352},
archivePrefix = {arXiv},
       eprint = {0709.1046},
 primaryClass = {astro-ph},
       adsurl = {https://ui.adsabs.harvard.edu/abs/2008ApJ...673..544Y}
}

@ARTICLE{2007A&A...474..239J,
       author = {{Jouve}, L. and {Brun}, A.~S.},
        title = "{On the role of meridional flows in flux transport dynamo models}",
      journal = {\aap},
         year = 2007,
        month = oct,
       volume = {474},
       number = {1},
        pages = {239-250},
          doi = {10.1051/0004-6361:20077070},
archivePrefix = {arXiv},
       eprint = {0712.3200},
 primaryClass = {astro-ph},
       adsurl = {https://ui.adsabs.harvard.edu/abs/2007A&A...474..239J}
}

@ARTICLE{2013A&A...555A.136V,
       author = {{Verma}, M. and {Steffen}, M. and {Denker}, C.},
        title = "{Evaluating local correlation tracking using CO5BOLD simulations of solar granulation}",
      journal = {\aap},
         year = 2013,
        month = jul,
       volume = {555},
          eid = {A136},
        pages = {A136},
          doi = {10.1051/0004-6361/201321628},
archivePrefix = {arXiv},
       eprint = {1305.6033},
 primaryClass = {astro-ph.SR},
       adsurl = {https://ui.adsabs.harvard.edu/abs/2013A&A...555A.136V}
}

@ARTICLE{1999ApJ...512..458B,
       author = {{Basu}, Sarbani and {Antia}, H.~M. and {Tripathy}, S.~C.},
        title = "{Ring Diagram Analysis of Near-Surface Flows in the Sun}",
      journal = {\apj},
         year = 1999,
        month = feb,
       volume = {512},
       number = {1},
        pages = {458-470},
          doi = {10.1086/306765},
archivePrefix = {arXiv},
       eprint = {astro-ph/9809309},
 primaryClass = {astro-ph},
       adsurl = {https://ui.adsabs.harvard.edu/abs/1999ApJ...512..458B}
}

@ARTICLE{2010A&A...520A..29W,
       author = {{W{\"o}hl}, H. and {Braj{\v{s}}a}, R. and {Hanslmeier}, A. and {Gissot}, S.~F.},
        title = "{A precise measurement of the solar differential rotation by tracing small bright coronal structures in SOHO-EIT images. Results and comparisons for the period 1998-2006}",
      journal = {\aap},
         year = 2010,
        month = sep,
       volume = {520},
          eid = {A29},
        pages = {A29},
          doi = {10.1051/0004-6361/200913081},
       adsurl = {https://ui.adsabs.harvard.edu/abs/2010A&A...520A..29W}
}

@ARTICLE{2009MNRAS.400L..34C,
       author = {{Chandra}, Satish and {Vats}, Hari Om and {Iyer}, K.~N.},
        title = "{Differential coronal rotation using radio images at 17GHz}",
      journal = {\mnras},
         year = 2009,
        month = nov,
       volume = {400},
       number = {1},
        pages = {L34-L37},
          doi = {10.1111/j.1745-3933.2009.00757.x},
archivePrefix = {arXiv},
       eprint = {0909.2786},
 primaryClass = {astro-ph.SR},
       adsurl = {https://ui.adsabs.harvard.edu/abs/2009MNRAS.400L..34C}
}

@ARTICLE{2005LRSP....2....5S,
       author = {{Sheeley}, Neil R., Jr.},
        title = "{Surface Evolution of the Sun's Magnetic Field: A Historical Review of the Flux-Transport Mechanism}",
      journal = {Living Reviews in Solar Physics},
         year = 2005,
        month = oct,
       volume = {2},
       number = {1},
          eid = {5},
        pages = {5},
          doi = {10.12942/lrsp-2005-5},
       adsurl = {https://ui.adsabs.harvard.edu/abs/2005LRSP....2....5S}
}

@ARTICLE{2005LRSP....2....2C,
       author = {{Charbonneau}, Paul},
        title = "{Dynamo Models of the Solar Cycle}",
      journal = {Living Reviews in Solar Physics},
         year = 2005,
        month = jun,
       volume = {2},
       number = {1},
          eid = {2},
        pages = {2},
          doi = {10.12942/lrsp-2005-2},
       adsurl = {https://ui.adsabs.harvard.edu/abs/2005LRSP....2....2C}
}

@ARTICLE{2018SoPh..293..145K,
       author = {{Komm}, R. and {Howe}, R. and {Hill}, F.},
        title = "{Subsurface Zonal and Meridional Flow During Cycles 23 and 24}",
      journal = {\solphys},
         year = 2018,
        month = oct,
       volume = {293},
       number = {10},
          eid = {145},
        pages = {145},
          doi = {10.1007/s11207-018-1365-7},
       adsurl = {https://ui.adsabs.harvard.edu/abs/2018SoPh..293..145K}
}

@ARTICLE{2010ApJ...725..658U,
       author = {{Ulrich}, Roger K.},
        title = "{Solar Meridional Circulation from Doppler Shifts of the Fe I Line at 5250 {\r{A}} as Measured by the 150-foot Solar Tower Telescope at the Mt. Wilson Observatory}",
      journal = {\apj},
         year = 2010,
        month = dec,
       volume = {725},
       number = {1},
        pages = {658-669},
          doi = {10.1088/0004-637X/725/1/658},
archivePrefix = {arXiv},
       eprint = {1010.0487},
 primaryClass = {astro-ph.SR},
       adsurl = {https://ui.adsabs.harvard.edu/abs/2010ApJ...725..658U}
}

@ARTICLE{1980ApJ...239L..33H,
       author = {{Howard}, R. and {Labonte}, B.~J.},
        title = "{The sun is observed to be a torsional oscillator with a period of 11 years}",
      journal = {\apjl},
         year = 1980,
        month = jul,
       volume = {239},
        pages = {L33-L36},
          doi = {10.1086/183286},
       adsurl = {https://ui.adsabs.harvard.edu/abs/1980ApJ...239L..33H}
}

@ARTICLE{1988ApJ...333..427N,
       author = {{November}, Laurence J. and {Simon}, George W.},
        title = "{Precise Proper-Motion Measurement of Solar Granulation}",
      journal = {\apj},
         year = 1988,
        month = oct,
       volume = {333},
        pages = {427},
          doi = {10.1086/166758},
       adsurl = {https://ui.adsabs.harvard.edu/abs/1988ApJ...333..427N}
}

@ARTICLE{2011A&A...529A.153V,
       author = {{Verma}, M. and {Denker}, C.},
        title = "{Horizontal flow fields observed in Hinode G-band images. I. Methods}",
      journal = {\aap},
         year = 2011,
        month = may,
       volume = {529},
          eid = {A153},
        pages = {A153},
          doi = {10.1051/0004-6361/201016358},
archivePrefix = {arXiv},
       eprint = {1103.2622},
 primaryClass = {astro-ph.SR},
       adsurl = {https://ui.adsabs.harvard.edu/abs/2011A&A...529A.153V}
}

@ARTICLE{2017A&A...604A..11A,
       author = {{Asensio Ramos}, A. and {Requerey}, I.~S. and {Vitas}, N.},
        title = "{DeepVel: Deep learning for the estimation of horizontal velocities at the solar surface}",
      journal = {\aap},
         year = 2017,
        month = jul,
       volume = {604},
          eid = {A11},
        pages = {A11},
          doi = {10.1051/0004-6361/201730783},
archivePrefix = {arXiv},
       eprint = {1703.05128},
 primaryClass = {astro-ph.SR},
       adsurl = {https://ui.adsabs.harvard.edu/abs/2017A&A...604A..11A}
}

@ARTICLE{2000SoPh..191...47B,
       author = {{Beck}, John G.},
        title = "{A comparison of differential rotation measurements - (Invited Review)}",
      journal = {\solphys},
         year = 2000,
        month = jan,
       volume = {191},
       number = {1},
        pages = {47-70},
          doi = {10.1023/A:1005226402796},
       adsurl = {https://ui.adsabs.harvard.edu/abs/2000SoPh..191...47B}
}

@ARTICLE{2013ApJ...774L..29Z,
       author = {{Zhao}, Junwei and {Bogart}, R.~S. and {Kosovichev}, A.~G. and {Duvall}, T.~L., Jr. and {Hartlep}, Thomas},
        title = "{Detection of Equatorward Meridional Flow and Evidence of Double-cell Meridional Circulation inside the Sun}",
      journal = {\apjl},
         year = 2013,
        month = sep,
       volume = {774},
       number = {2},
          eid = {L29},
        pages = {L29},
          doi = {10.1088/2041-8205/774/2/L29},
archivePrefix = {arXiv},
       eprint = {1307.8422},
 primaryClass = {astro-ph.SR},
       adsurl = {https://ui.adsabs.harvard.edu/abs/2013ApJ...774L..29Z}
}

@ARTICLE{2020Sci...368.1469G,
       author = {{Gizon}, Laurent and {Cameron}, Robert H. and {Pourabdian}, Majid and {Liang}, Zhi-Chao and {Fournier}, Damien and {Birch}, Aaron C. and {Hanson}, Chris S.},
        title = "{Meridional flow in the Sun{\textquoteright}s convection zone is a single cell in each hemisphere}",
      journal = {Science},
         year = 2020,
        month = jun,
       volume = {368},
       number = {6498},
        pages = {1469-1472},
          doi = {10.1126/science.aaz7119},
       adsurl = {https://ui.adsabs.harvard.edu/abs/2020Sci...368.1469G}
}

@ARTICLE{2010Sci...327.1350H,
       author = {{Hathaway}, David H. and {Rightmire}, Lisa},
        title = "{Variations in the Sun{\textquoteright}s Meridional Flow over a Solar Cycle}",
      journal = {Science},
         year = 2010,
        month = mar,
       volume = {327},
       number = {5971},
        pages = {1350},
          doi = {10.1126/science.1181990},
       adsurl = {https://ui.adsabs.harvard.edu/abs/2010Sci...327.1350H}
}

@ARTICLE{2014ApJ...780....5U,
       author = {{Upton}, Lisa and {Hathaway}, David H.},
        title = "{Predicting the Sun's Polar Magnetic Fields with a Surface Flux Transport Model}",
      journal = {\apj},
         year = 2014,
        month = jan,
       volume = {780},
       number = {1},
          eid = {5},
        pages = {5},
          doi = {10.1088/0004-637X/780/1/5},
archivePrefix = {arXiv},
       eprint = {1311.0844},
 primaryClass = {astro-ph.SR},
       adsurl = {https://ui.adsabs.harvard.edu/abs/2014ApJ...780....5U}
}

@ARTICLE{2011ApJ...729...80H,
       author = {{Hathaway}, David H. and {Rightmire}, Lisa},
        title = "{Variations in the Axisymmetric Transport of Magnetic Elements on the Sun: 1996-2010}",
      journal = {\apj},
         year = 2011,
        month = mar,
       volume = {729},
       number = {2},
          eid = {80},
        pages = {80},
          doi = {10.1088/0004-637X/729/2/80},
archivePrefix = {arXiv},
       eprint = {1010.1242},
 primaryClass = {astro-ph.SR},
       adsurl = {https://ui.adsabs.harvard.edu/abs/2011ApJ...729...80H}
}

@ARTICLE{2014JGRA..119.3316H,
       author = {{Hathaway}, D.~H. and {Upton}, L.},
        title = "{The solar meridional circulation and sunspot cycle variability}",
      journal = {Journal of Geophysical Research (Space Physics)},
         year = 2014,
        month = may,
       volume = {119},
       number = {5},
        pages = {3316-3324},
          doi = {10.1002/2013JA019432},
archivePrefix = {arXiv},
       eprint = {1404.5893},
 primaryClass = {astro-ph.SR},
       adsurl = {https://ui.adsabs.harvard.edu/abs/2014JGRA..119.3316H}
}

@ARTICLE{2014ApJ...792..142U,
       author = {{Upton}, Lisa and {Hathaway}, David H.},
        title = "{Effects of Meridional Flow Variations on Solar Cycles 23 and 24}",
      journal = apj,
         year = 2014,
        month = sep,
       volume = {792},
       number = {2},
          eid = {142},
        pages = {142},
          doi = {10.1088/0004-637X/792/2/142},
archivePrefix = {arXiv},
       eprint = {1408.0035},
 primaryClass = {astro-ph.SR},
       adsurl = {https://ui.adsabs.harvard.edu/abs/2014ApJ...792..142U}
}

@ARTICLE{2004ApJ...612.1181L,
       author = {{Longcope}, D.~W.},
        title = "{Inferring a Photospheric Velocity Field from a Sequence of Vector Magnetograms: The Minimum Energy Fit}",
      journal = {\apj},
         year = 2004,
        month = sep,
       volume = {612},
       number = {2},
        pages = {1181-1192},
          doi = {10.1086/422579},
       adsurl = {https://ui.adsabs.harvard.edu/abs/2004ApJ...612.1181L}
}

@ARTICLE{2013A&A...550A..19R,
       author = {{Ravindra}, B. and {Priya}, T.~G. and {Amareswari}, K. and {Priyal}, M. and {Nazia}, A.~A. and {Banerjee}, D.},
        title = "{Digitized archive of the Kodaikanal images: Representative results of solar cycle variation from sunspot area determination}",
      journal = {\aap},
         year = 2013,
        month = feb,
       volume = {550},
          eid = {A19},
        pages = {A19},
          doi = {10.1051/0004-6361/201220416},
archivePrefix = {arXiv},
       eprint = {1212.4776},
 primaryClass = {astro-ph.SR},
       adsurl = {https://ui.adsabs.harvard.edu/abs/2013A&A...550A..19R}
}

@ARTICLE{2017A&A...601A.106M,
       author = {{Mandal}, Sudip and {Hegde}, Manjunath and {Samanta}, Tanmoy and {Hazra}, Gopal and {Banerjee}, Dipankar and {Ravindra}, B.},
        title = "{Kodaikanal digitized white-light data archive (1921-2011): Analysis of various solar cycle features}",
      journal = {\aap},
         year = 2017,
        month = may,
       volume = {601},
          eid = {A106},
        pages = {A106},
          doi = {10.1051/0004-6361/201628651},
archivePrefix = {arXiv},
       eprint = {1608.04665},
 primaryClass = {astro-ph.SR},
       adsurl = {https://ui.adsabs.harvard.edu/abs/2017A&A...601A.106M}
}

@ARTICLE{2025SoPh..300..149K,
       author = {{Komm}, R.~W. and {Howe}, R.},
        title = "{Solar-Cycle Variation of Large-Scale Flows in the Near-Surface Shear Layer from SC 23 to SC 26}",
      journal = {\solphys},
         year = 2025,
        month = nov,
       volume = {300},
          eid = {149},
        pages = {149},
          doi = {10.1007/s11207-025-02566-1},
       adsurl = {https://ui.adsabs.harvard.edu/abs/2025SoPh..300..149K}
}

@ARTICLE{1993Natur.362..430D,
       author = {{Duvall}, Jr., T.~L. and {Jefferies}, S.~M. and {Harvey}, J.~W. and {Pomerantz}, M.~A.},
        title = "{Time-distance helioseismology}",
      journal = {Nature},
         year = 1993,
        month = mar,
       volume = {362},
       number = {6419},
        pages = {430-432},
          doi = {10.1038/362430a0},
       adsurl = {https://ui.adsabs.harvard.edu/abs/1993Natur.362..430D}
}

@ARTICLE{2005MmSAI..76.1018O,
       author = {{Ortiz}, A. and {Rast}, M.},
        title = "{How good is the Ca II K as a proxy for the magnetic flux?}",
      journal = {Memorie della Societa Astronomica Italiana Supplementi},
         year = 2005,
       volume = {76},
        pages = {1018},
       adsurl = {https://ui.adsabs.harvard.edu/abs/2005MmSAI..76.1018O}
}

@ARTICLE{2014SoPh..289..137P,
       author = {{Priyal}, Muthu and {Singh}, Jagdev and {Ravindra}, B. and {Priya}, T.~G. and {Amareswari}, K.},
        title = "{Long Term Variations in Chromospheric Features from Ca-K Images at Kodaikanal}",
      journal = {Solar Physics},
         year = 2014,
        month = jan,
       volume = {289},
       number = {1},
        pages = {137-152},
          doi = {10.1007/s11207-013-0315-7},
       adsurl = {https://ui.adsabs.harvard.edu/abs/2014SoPh..289..137P}
}

@ARTICLE{2017SoPh..292...85P,
       author = {{Priyal}, Muthu and {Singh}, Jagdev and {Belur}, Ravindra and {Rathina}, Somasundaram Kameshwaran},
        title = "{Long-term Variations in the Intensity of Plages and Networks as Observed in Kodaikanal Ca-K Digitized Data}",
      journal = {Solar Physics},
         year = 2017,
        month = jul,
       volume = {292},
       number = {7},
          eid = {85},
        pages = {85},
          doi = {10.1007/s11207-017-1106-3},
       adsurl = {https://ui.adsabs.harvard.edu/abs/2017SoPh..292...85P}
}

@ARTICLE{2024ApJ...961...40M,
       author = {{Mishra}, Dibya Kirti and {Routh}, Srinjana and {Jha}, Bibhuti Kumar and {Chatzistergos}, Theodosios and {Basu}, Judhajeet and {Chatterjee}, Subhamoy and {Banerjee}, Dipankar and {Ermolli}, Ilaria},
        title = "{Differential Rotation of the Solar Chromosphere: A Century-long Perspective from Kodaikanal Solar Observatory Ca II K Data}",
      journal = {The Astrophysical Journal},
         year = 2024,
        month = feb,
       volume = {961},
          eid = {40},
        pages = {40},
          doi = {10.3847/1538-4357/ad1188},
archivePrefix = {arXiv},
       eprint = {2311.18800},
 primaryClass = {astro-ph.SR},
       adsurl = {https://ui.adsabs.harvard.edu/abs/2024ApJ...961...40M}
}

@ARTICLE{1990AJ....100...32B,
       author = {{Beers}, Timothy C. and {Flynn}, Kevin and {Gebhardt}, Karl},
        title = "{Measures of Location and Scale for Velocities in Clusters of Galaxies---A Robust Approach}",
      journal = {The Astronomical Journal},
         year = 1990,
        month = jul,
       volume = {100},
        pages = {32},
          doi = {10.1086/115487},
       adsurl = {https://ui.adsabs.harvard.edu/abs/1990AJ....100...32B}
}

@ARTICLE{1990SoPh..130..295H,
       author = {{Howard}, Robert F. and {Harvey}, John W. and {Forgach}, Sandor},
        title = "{Solar surface velocity fields determined from small magnetic features}",
      journal = {Solar Physics},
         year = 1990,
        month = dec,
       volume = {130},
          eid = {2},
        pages = {295},
          doi = {10.1007/BF00156795},
       adsurl = {https://ui.adsabs.harvard.edu/abs/1990SoPh..130..295H}
}

@ARTICLE{1951MNRAS..111..413N,
       author = {{Newton}, H. W. and {Nunn}, M. L.},
        title = "{The Sun's rotation derived from sunspots 1934-1944 and additional results}",
      journal = {Monthly Notices of the Royal Astronomical Society},
         year = 1951,
        month = aug,
       volume = {111},
          eid = {4},
        pages = {413},
          doi = {10.1093/mnras/111.4.413},
       adsurl = {https://ui.adsabs.harvard.edu/abs/1951MNRAS..111..413N}
}

@ARTICLE{2020LRSP...17....4C,
       author = {{Charbonneau}, Paul},
        title = "{Dynamo models of the solar cycle}",
      journal = {Living Reviews in Solar Physics},
         year = 2020,
        month = dec,
       volume = {17},
       number = {1},
          eid = {4},
        pages = {4},
          doi = {10.1007/s41116-020-00025-6},
       adsurl = {https://ui.adsabs.harvard.edu/abs/2020LRSP...17....4C}
}

@ARTICLE{2023ApJ...950...63M,
       author = {{Mahajan}, Sushant S. and {Sun}, Xudong and {Zhao}, Junwei},
        title = "{Removal of Active Region Inflows Reveals a Weak Solar Cycle Scale Trend in the Near-surface Meridional Flow}",
      journal = {\apj},
         year = 2023,
        month = jun,
       volume = {950},
       number = {1},
          eid = {63},
        pages = {63},
          doi = {10.3847/1538-4357/acc839},
archivePrefix = {arXiv},
       eprint = {2304.02158},
 primaryClass = {astro-ph.SR},
       adsurl = {https://ui.adsabs.harvard.edu/abs/2023ApJ...950...63M}
}

@ARTICLE{2003SoPh..212...23J,
       author = {{Javaraiah}, J.},
        title = "{Long-Term Variations in the Solar Differential Rotation}",
      journal = {Solar Physics},
         year = 2003,
        month = jan,
       volume = {212},
       number = {1},
        pages = {23--49},
          doi = {10.1023/A:1022912430585},
       adsurl = {https://ui.adsabs.harvard.edu/abs/2003SoPh..212...23J}
}

@ARTICLE{2017A&A...606A..72P,
       author = {{Poljan{\v{c}}i{\'c} Beljan}, I. and {Jurdana-{\v{S}}epi{\'c}}, R. and {Braj{\v{s}}a}, R. and {Sudar}, D. and {Ru{\v{z}}djak}, D. and {Hr{\v{z}}ina}, D. and {P{\"o}tzi}, W. and {Hanslmeier}, A. and {Veronig}, A. and {Skoki{\'c}}, I. and {W{\"o}hl}, H.},
        title = "{Solar differential rotation in the period 1964--2016 determined by the Kanzelh{\"o}he data set}",
      journal = {Astronomy \& Astrophysics},
         year = 2017,
        month = oct,
       volume = {606},
          eid = {A72},
        pages = {A72},
          doi = {10.1051/0004-6361/201731047},
archivePrefix = {arXiv},
       eprint = {1707.07886},
 primaryClass = {astro-ph.SR},
       adsurl = {https://ui.adsabs.harvard.edu/abs/2017A&A...606A..72P}
}

@ARTICLE{2024ApJ...972...46S,
       author = {{Shokri}, Zahra and {Alipour}, Nasibe and {Safari}, Hossein},
        title = "{Solar Rotation and Activity for Cycle 24 from SDO/AIA Observations}",
      journal = {The Astrophysical Journal},
         year = 2024,
        month = aug,
       volume = {972},
       number = {1},
          eid = {46},
        pages = {46},
          doi = {10.3847/1538-4357/ad58c0},
archivePrefix = {arXiv},
       eprint = {2407.17594},
 primaryClass = {astro-ph.SR},
       adsurl = {https://ui.adsabs.harvard.edu/abs/2024ApJ...972...46S}
}

@ARTICLE{2018SoPh..293...29L,
       author = {{Larson}, Timothy P. and {Schou}, Jesper},
        title = "{Global-Mode Analysis of Full-Disk Data from the Michelson Doppler Imager and the Helioseismic and Magnetic Imager}",
      journal = {Solar Physics},
         year = 2018,
        month = jan,
       volume = {293},
       number = {2},
          eid = {29},
        pages = {29},
          doi = {10.1007/s11207-017-1201-5},
       adsurl = {https://ui.adsabs.harvard.edu/abs/2018SoPh..293...29L}
}

@ARTICLE{2026arXiv260819438R,
       author = {{Rabello Soares}, M.~Cristina and {Basu}, Sarbani and {Bogart}, Richard S.},
        title = "{Correlations with Magnetic Activity in the Solar Near-Surface Shear Layer. I. Rotation}",
      journal = {The Astrophysical Journal},
         year = 2026,
archivePrefix = {arXiv},
       eprint = {2608.19438},
 primaryClass = {astro-ph.SR},
          doi = {10.3847/1538-4357/ae985b},
         note = {Accepted; DOI listed on arXiv may not yet resolve},
       adsurl = {https://ui.adsabs.harvard.edu/abs/2026arXiv260819438R}
}
\bibliographystyle{aasjournalv7}



\end{document}